# JWST early Universe observations and ΛCDM cosmology

Rajendra P. Gupta[i]

*Department of Physics, University of Ottawa, Ottawa, Canada*
*Department of Physics, Carleton University, Ottawa, Canada*

**ABSTRACT**

Deep space observations of the James Webb Space Telescope (JWST) have revealed that the structure and masses of very early Universe galaxies at high redshifts ($z$~15), existing at ~0.3 Gyr after the BigBang, may be as evolved as the galaxies in existence for ~10 Gyr. The JWST findings are thus in strong tension with the ΛCDM cosmological model. While tired light (TL) models have been shown to comply with the JWST angular galaxy size data, they cannot satisfactorily explain isotropy of the cosmic microwave background (CMB) observations or fit the supernovae distance modulus vs. redshift data well. We have developed hybrid models that include the tired light concept in the expanding universe. The hybrid ΛCDM model fits the supernovae type 1a data well but not the JWST observations. We present a model with covarying coupling constants (CCC), starting from the modified FLRW metric and resulting Einstein and Friedmann equations, and a CCC+TL hybrid model. They fit the Pantheon+ data admirably, and the CCC+TL model is compliant with the JWST observations. It stretches the age of the universe to 26.7 Gyr with 5.8 Gyr at $z = 10$ and 3.5 Gyr at $z = 20$, giving enough time to form massive galaxies. It thus resolves the 'impossible early galaxy' problem without requiring the existence of primordial black hole seeds or modified power spectrum, rapid formation of massive population III stars, and super Eddington accretion rates. One could infer the CCC model as an extension of the ΛCDM model with a dynamic cosmological constant.

**Key words**: Cosmology, galaxies, (cosmology:) early Universe < Cosmology, galaxies: high-redshift < galaxies

## I. INTRODUCTION

Observations with the James Webb Space Telescope (JWST) have revealed the existence of massive, bright galaxies in the very young universe of age ~500 million years (e.g., Naidu et al. 2022a, 2022b; Labbe et al. 2023; Curtis-Lake et al. 2023; Hainline et al. 2023; Robertson et al. 2023), which is a small fraction of its current age estimate of 13.7 billion years according to the standard ΛCDM model. Angular diameters of many such galaxies are an order of magnitude smaller than expected from the ΛCDM model (e.g., Adams et al. 2022, Atek et al. 2022, Chen et al. 2022, Donnan et al. 2022, Finkelstein et al. 2022, Naidu et al. 2022a, 2022b, Ono et al. 2022, Tacchella et al. 2022, Wu et al. 2022, Yang et al. 2022, Austin et al. 2023, Baggen et al. 2023). Astronomers first identified such an 'impossible early galaxy' problem from observations with the Hubble Space Telescope (HST) at high redshifts $z$~10 (Melia 2014, 2020, 2023). Dekel et al. (2023) stated 'JWST observations reveal a surprising excess of luminous galaxies at $z$~10 '. According to Boyett et al. (2023) 'JWST observations confirm the existence of galaxies as early as 300 Myr and at a higher number density than expected based on galaxy formation models and HST observations'. Looser et al. (2023) observed the existence of a quiescent galaxy when the Universe was only 700 Myr old (see also Long et al. 2023). Bunker et al. (2023) wrote 'Our NIRSpec spectroscopy confirms that GN-z11 is a remarkable galaxy with extreme properties seen 430 Myr after the Big Bang' (see also Tacchella et al. 2023); NIRSpec refers to 'near infrared spectrograph' and GN-z11 nomenclature means Good North survey of galaxy with $z \cong 11$. Following spectroscopic confirmation of several photometric redshifts from JWST early galaxy observations, Haro et al. (2023) stated 'our results solidifies photometric evidence for a high space density of bright galaxies at $z > 8$ compared to theoretical model predictions'. Related to the massive quasar analysis observed at $z > 6$, Eilers et al. (2023) wrote '..this quasar hosts a ten billion solar mass black hole less than 1 Gyr after the Big Bang, which is challenging to explain with current black hole formation models'. 'And the findings have been dazzling astronomers, revealing that stars and galaxies were forming and evolving much earlier than anyone had expected' as expressed by Alexandra Witze (2023) in a recent news article in *Nature*. How come the galaxies in the very early universe were as evolved as those with billions of years of evolution, some as early as less than ~300 million years after the big bang? These astonishing observations make the problem even more acute to resolve by tweaking well-established galaxy formation and cosmological models developed to satisfy lower redshifts observations (e.g., Haslbauer et al. 2022; Inayoshi et al. 2022; Kannan et al. 2022; Keller et al. 2022; Regan 2022; Yajima et al. 2022;

Atek et al. 2023, Mason et al. 2023; Mirocha & Furlanetto 2023; Whitler et al. 2023a, 2023b; McCaffrey et al. 2023). Attempts have been made to compress time for the formation of population III stars and galaxies more and more, such as by considering the presence of primordial massive black hole seeds, and super-Eddington accretion rates in the early Universe (Ellis 2022, Bastian et al. 2023, Brummel-Smith 2023, Chantavat et al. 2023, Dolgov 2023, Larson et al. 2023, Maiolino et al. 2023). While analysing GN-Z11 JWST-NIRSpec data, Maiolino et al. (2023) concluded that the black hole seed of this exceptionally luminous galaxy at $z = 10.6$ must be accreting at an episodic rate of about five time the Eddington rate for 100 million years (since $z \sim 12 - 15$) and is challenging for theoretical models (see also Schneider et al. 2023). Chen et al. (2023) considered the presence of massive dark matter halos (see also Mauerhofer and Dayal). As discussed elegantly by Melisa (2023), both are considered unrealistic based on the zero angular momentum argument and observations showing most of the distant quasars accreting below or at the Eddington limit (Melia 2023).

As stated by Wang and Liu (2023) 'JWST high redshift galaxy observations have a strong tension with Planck CMB measurements'. They could not resolve the tension using alternative cosmological models either, including dark matter-baryon interaction, f(R) gravity, and dynamical dark energy (see also Santini et al. 2023). Parashari and Laha (2023) have shown that a blue-tilted power spectrum could potentially alleviate the tension with a low to moderate star formation efficiency. Studying the Balmer breaks at highest redshift galaxies, Steinhardt et al (2023) proposed it as a test of the ΛCDM model. They emphasized that the existence of stronger Balmer breaks out to $z \gtrsim 11$, will either demonstrate the early galaxy formation templates are invalid at high redshift or imply new physics beyond the 'vanilla' ΛCDM model. Lovyagin et al. (2022) have shown that the impossible early galaxy problem can be resolved amicably with the tired light theory in a steady-state Universe, first advocated by Zwicky (1929) to explain the early redshift observations by Hubble (1929). They have succinctly reviewed the current status of the problem and suggested some solutions. However, the tired light model cannot explain the extreme directional uniformity of the observed cosmic microwave background (CMB) radiation (Penzias and Wilson 1965). Additionally, the standard tired light model does not fit the supernovae type 1a data except at very low redshifts. An expanding Universe model can easily account for the observed redshift of distant galaxies and the CMB isotropy but has problems with the early Universe observations. However, the two approaches are not mutually exclusive. Gupta (2018a, 2018b) has contemplated the existence of tired light in an expanding Universe. Our main objective in this paper is to explore if hybrid tired light and expanding Universe models can resolve the impossible early galaxy problem without conflicting with the relatively low redshift $z \leq 2.5$ data from supernovae type 1a (SNe Ia) observations in distant galaxies such as the Pantheon+ data (Scolnic et al. 2022). Since the distance traveled by light reaching us is the same irrespective of whether in the tired light scenario or the expanding universe, it constrains the parameters in them; *one does not need an extra parameter to include tired light in an expanding Universe model*. We will use two expanding Universe models to develop hybrid models, compare them, and see how they fit the SNe Ia and the galaxy size data from HST and JWST observations. The first hybrid model combines the ΛCDM model with the tired light model and is named the ΛCDM+TL model. The second hybrid model incorporates a new model derived from covarying coupling constant (CCC) approach (Gupta 2022). We start by defining an appropriate metric for the CCC approach and develop Einstein and Friedmann equations for the new CCC cosmological model; the hybrid model is dubbed the CCC+TL model.

Section II of this paper develops the theory of the expanding Universe models and their hybrid counterparts, especially for the CCC model, and shows how it relates to the standard ΛCDM model. In Section III, we attempt to fit the Pantheon+ data (Scolnic et al. 2022, Brout et al. 2022) for the models developed in Section II to qualify them for testing further with the HST and JWST data. We discuss our findings in Section IV and present the conclusions in Section V.

## II. THEORY

*Basics:* We will derive the Friedmann equations starting from the Einstein equations and see how they are modified when $G$ and $c$ are varying such that $G \sim c^3$ and the speed of light is used to measure distances (Gupta 2022). The basic Einstein equations (without the cosmological constant) are

$$G_{\mu\nu} = \frac{8\pi G(t)}{c(t)^4} T_{\mu\nu}, \qquad (1)$$

where $G_{\mu\nu}$ is the Einstein tensor and $T_{\mu\nu}$ is the stress-energy tensor. The FLRW metric takes the form

$$ds^2 = c^2 dt^2 f(t)^2 - a(t)^2 f(t)^2 \left( \frac{dr^2}{1-kr^2} + r^2 (d\theta^2 + \sin^2\theta \, d\phi^2) \right), \quad (2)$$

where $f(t) = \exp(\alpha(t - t_0))$ defines the variation of $c$ and the distances are measured using the speed of light; $t_0$ is the current time when $f(t_0) = 1$, $a$ is the scale factor, and $k$ $(-1, 0, 1)$ determines the geometry of the space. The most significant difference from the standard FLRW metric is that the $g_{00}$ metric coefficient is time dependent (Gomide & Oehara 1981). Since $\dot{f}/f = \alpha$ and $\ddot{f}/f = \alpha^2$, and $G_{\mu\nu}$ is a diagonal matrix, the essential components of the Einstein tensor are

$$G_{00} = 3c^{-2}\left(\frac{kc^2}{a^2} + \alpha^2 + 2\alpha\frac{\dot{a}}{a} + \left(\frac{\dot{a}}{a}\right)^2\right), \text{ and} \tag{3}$$

$$G_{11} = -\frac{a^2}{c^2(1-kr^2)}\left(\frac{kc^2}{a^2} + \alpha^2 + 4\alpha\frac{\dot{a}}{a} + \left(\frac{\dot{a}}{a}\right)^2 + 2\frac{\ddot{a}}{a}\right). \tag{4}$$

The stress-energy tensor for a homogeneous and isotropic universe considered a perfect fluid, has the form

$$T_{\mu\nu} = c^{-2}(\varepsilon + p)u_\mu u_\nu - p g_{\mu\nu}. \tag{5}$$

Here $\varepsilon$ is the energy density, $p$ is the pressure, $g_{\mu\nu}$ are the metric elements of the FLRW metric, and the four-velocity $u_\mu$ are related through $g^{\mu\nu}u_\mu u_\nu = c^2$. For a comoving observer, $u_\mu = (c, 0, 0, 0)$.

Since $G_{\mu\nu}$ satisfies the contracted Bianchi identities and $T_{\mu\nu}$ obeys the local conservation laws, we have

$$\nabla^\mu G_{\mu\nu} = 0, \text{ i.e. } \nabla^\mu \left(\frac{8\pi G(t)}{c(t)^4} T_{\mu\nu}\right) = 0, \text{ and } \nabla^\mu T_{\mu\nu} = 0. \tag{6}$$

Since $T_{\mu\nu}$ is also a diagonal matrix, and spatial components are all equal, we need to consider only $T_{00}$ and $T_{11}$. Now $g_{00} = f^2$, $g^{00} = f^{-2}$. Therefore, $g^{\mu\nu}u_\mu u_\nu = c^2 \Rightarrow u_0 u_0 = f^2 c^2$, and

$$T_{00} = (\varepsilon + p)f^2 - pf^2 = \varepsilon f^2, \tag{7}$$

$$T_{11} = \frac{pf^2 a^2}{1-kr^2}. \tag{8}$$

Now, $\varepsilon$ and $p$ both are dimensionally the same. Also, $\varepsilon = \rho c^2 \sim mc^2/r^3 \sim f^2/f^3 \sim f^{-1}$ when $m$ does not vary with time. It means $T_{00}$ and $T_{11}$ both scale as $f$. Since $c \sim f$ and $G \sim f^3$, $8\pi G T_{\mu\nu}/c^4 \sim f^0$. Therefore, $\nabla^\mu(8\pi G T_{\mu\nu}/c^4) = 0$ when $\nabla^\mu T_{\mu\nu} = 0$ with the standard definition of $T_{\mu\nu}$ without the function $f^2$. Einstein equations corresponding to the metric of Eq. (2) may now be written

$$\frac{kc^2}{a^2} + \alpha^2 + 2\alpha\frac{\dot{a}}{a} + \left(\frac{\dot{a}}{a}\right)^2 = \frac{8\pi G}{3c^2}\varepsilon, \text{ and} \tag{9}$$

$$\frac{kc^2}{a^2} + \alpha^2 + 4\alpha\frac{\dot{a}}{a} + \left(\frac{\dot{a}}{a}\right)^2 + 2\frac{\ddot{a}}{a} = -\frac{8\pi G}{c^2}p. \tag{10}$$

Rearranging them leads to the Friedmann equations

$$\left(\frac{\dot{a}}{a}\right)^2 = \frac{8\pi G}{3c^2}\varepsilon - \frac{kc^2}{a^2} - \left(\alpha^2 + 2\alpha\frac{\dot{a}}{a}\right), \text{ and} \tag{11}$$

$$\frac{\ddot{a}}{a} = -\frac{4\pi G}{3c^2}(\varepsilon + 3p) - \alpha\left(\frac{\dot{a}}{a}\right). \tag{12}$$

From these equations, we obtain the continuity equation

$$\dot{\varepsilon} + 3\frac{\dot{a}}{a}(\varepsilon + p) = -\alpha(\varepsilon + 3p). \tag{13}$$

The solution of this equation for the matter-dominant ($p = 0$) and radiation-dominant ($p = \varepsilon/3$) epochs of the universe are, respectively,

$$\varepsilon = \varepsilon_0 a^{-3} \exp(-\alpha(t - t_0)) = \varepsilon_0 a^{-3} f^{-1}, \text{ and } \varepsilon = \varepsilon_0 a^{-4} \exp(-2\alpha(t - t_0)) = \varepsilon_0 a^{-4} f^{-2}. \tag{14}$$

We will label these equations as representing the CCC (covarying coupling constants) universe. Comparing Eqs. (11) to (14) with corresponding equations for the ΛCDM model,

$$\left(\frac{\dot{a}}{a}\right)^2 = \frac{8\pi G}{3c^2}\varepsilon - \frac{kc^2}{a^2} + \frac{\Lambda}{3}, \tag{15}$$

$$\frac{\ddot{a}}{a} = -\frac{4\pi G}{3c^2}(\varepsilon + 3p) + \frac{\Lambda}{3}, \tag{16}$$

$$\dot{\varepsilon} + 3\frac{\dot{a}}{a}(\varepsilon + p) = 0, \tag{17}$$

$$\varepsilon = \varepsilon_0 a^{-3}, \text{ and } \varepsilon = \varepsilon_0 a^{-4}, \tag{18}$$

immediately reveals: (a) the cosmological constant Λ of the ΛCDM model is replaced with the constant $\alpha$ in the CCC model, (b) the continuity equation has an additional term involving $\alpha$ in the CCC model, and (c) energy density evolution has extra factors that must be considered for the CCC model.

*CCC Model*: Defining the Hubble expansion parameter as $H = \dot{a}/a$, we may write Eq. (11) for a flat universe ($k = 0$) as

$$(H + \alpha)^2 = \frac{8\pi G}{3c^2}\varepsilon \Rightarrow H_0 + \alpha = \sqrt{\frac{8\pi G}{3c^2}\varepsilon_0}. \tag{19}$$

In the matter-dominated universe of interest to us in this work, using Eq. (14), we may write

$$H = (H_0 + \alpha)a^{-(3/2)}f^{-(1/2)} - \alpha. \tag{20}$$

Since the observations are made using redshift $z$, we have to see how the scale factor $a$ relates to $z$ in the CCC model. Along the spatial geodesic ($\theta$ and $\phi$ constant) between the observer and the source at a fixed time $t$ using the modified FLRW metric (Eq. 2)

$$ds = a(t)f(t)dr. \tag{21}$$

Thus, the proper distance for commoving coordinate $r$ [recall that $a(t_0) = 1 = f(t_0)$]

$$d_p = a(t)f(t)\int_0^r dr = a(t)f(t)r \Rightarrow d_p(t_0) = r. \tag{22}$$

Since the light follows the null geodesic, Eq. (2) for a light emitted by a source at a time $t_e$ and detected by the observer at a time $t_0$ yields

$$c\int_{t_e}^{t_0} \frac{dt}{a(t)} = \int_0^r dr = r = d_p(t_0). \tag{23}$$

Thus, the expression for the proper distance is the same when using the standard FLRW metric. It can now be shown (Ryden 2017) that $a = 1/(1 + z)$, i.e., the same as for the ΛCDM model.

The next thing to consider is to transpose $f(t)$ to $f(z)$, as it is the latter that we will require in calculating the proper distance. We may write Eq. (20)

$$\frac{da}{dt} = -\alpha a + (H_0 + \alpha)f^{-(1/2)}a^{-(1/2)} \equiv -\alpha a + (H_0 + \alpha)a^{-(1/2)}\exp\left(-\frac{\alpha(t-t_0)}{2}\right). \tag{24}$$

Its analytic solution (using WolframAlpha), with the boundary condition $a = 1$ at $t = t_0$, is

$$a = \left(\frac{3}{2}\frac{(H_0+\alpha)}{\alpha}\exp\left(-\frac{\alpha(t-t_0)}{2}\right) + \left(1 - \frac{3}{2}\frac{(H_0+\alpha)}{\alpha}\right)\exp\left(-\frac{3\alpha(t-t_0)}{2}\right)\right)^{2/3}. \tag{25}$$

It can be written as a cubic equation

$$Ax^3 + Cx + D = 0, \text{ with } A = 1 - \frac{3}{2}\frac{(H_0+\alpha)}{\alpha} = 1 - C, D = -a^{3/2}, \text{ and } x = \exp\left(-\frac{\alpha(t-t_0)}{2}\right). \tag{26}$$

Its solution is

$$f^{-1/2} = x = \left(-\frac{D}{2A} + \left(\left(-\frac{D}{2A}\right)^2 + \left(\frac{C}{3A}\right)^3\right)^{1/2}\right)^{1/3} + \left(-\frac{D}{2A} - \left(\left(-\frac{D}{2A}\right)^2 + \left(\frac{C}{3A}\right)^3\right)^{1/2}\right)^{1/3} \tag{27}$$

Since the scale factor $a = 1/(1+z)$, we have $D = -[1/(1+z)]^{3/2}$. Thus, the above equation provides the function $f(z, H_0, \alpha)$.

Next, to determine the luminosity distance $d_L$ of an object, we need to understand how the photon energy flux evolves with the redshift in the CCC model. The photon energy is reduced by a factor $(1+z)$. The effect of the interval between photo arrival time on the flux is an additional factor $(1+z)$. Since in CCC, the distance scales as the speed of light, their evolution cancels out: two emitted photons separated in time by $\delta t_e$ are separated in space by a distance $\delta r_e = c_e \delta t_e$. This distance becomes $\delta r_0 = c_0 \delta t_e (1+z)$ and the corresponding time interval $\delta t_0 = c_0 \delta t_e (1+z)/c_0 = \delta t_e (1+z)$. Thus the luminosity distance scaling is the same as for the ΛCDM model, i.e. $D_L = d_p(t_0)(1+z)$, and the expression for the distance modulus $\mu$ also the same, viz.,

$$\mu = 5 \log_{10}(d_p/1\text{Mpc}) + 5 \log_{10}(1+z) + 25. \tag{28}$$

We will now focus on determining the expression for the proper distance $d_p$. Since $dt = dt \times da/da = da/\dot{a}$, we may write Eq. (23)

$$d_p(t_0) = c \int_{t_e}^{t_0} \frac{dt}{a(t)} = c \int_{a_e}^{1} \frac{da}{a\dot{a}} = c \int_{a_e}^{1} \frac{da}{a^2 H}, \tag{29}$$

and since $a = 1/(1+z), da = -dz/(1+z)^2 = -dz a^2$, we get, using Eq. (20)

$$d_p(t_0) = c \int_0^z \frac{dz}{H} = c \int_0^z \frac{dz}{(H_0+\alpha)(1+z)^{(3/2)} f(z)^{-(1/2)} - \alpha}. \tag{30}$$

*ΛCDM model:* The corresponding expression for the ΛCDM model (matter dominant, flat universe) is (Ryden 2017),

$$d_p(t_0) = \frac{c}{H_0} \int_0^z \frac{dz}{\sqrt{\Omega_{m,0}(1+z)^3 + 1 - \Omega_{m,0}}}, \tag{31}$$

where $\Omega_{m,0}$ is the relative matter energy density and $(1 - \Omega_{m,0})$ is the relative dark energy density. Notice that both the models have only two unknowns, i.e., $H_0$ and $\alpha$ for the CCC model, and $H_0$ and $\Omega_{m,0}$ for the ΛCDM model.

*Tired Light Model:* We would also like to consider the tired light model as it fits admirably well the high redshift galaxy size data (Lovyagin et al. 2022), especially from the James Webb Space Telescope (JWST). In this approach, one assumes that the photon energy loss, $hd\nu$, is proportional to its energy, $h\nu$, and the distance traveled, $dr$, i.e., $d\nu = K\nu dr$, with $K$ as the proportionality constant. Thus,

$$\frac{d\nu}{\nu} = Kdr \Rightarrow \int_{\nu_e}^{\nu_0} \frac{d\nu}{\nu} = K \int_{d_p}^{0} dr \Rightarrow \ln\left(\frac{\nu_0}{\nu_e}\right) = -Kd_p \Rightarrow \ln\left(\frac{\lambda_0}{\lambda_e}\right) = Kd_p. \tag{32}$$

Here $\nu_e$ and $\nu_0$ are the emitted and observed photon frequencies and $\lambda_e$ and $\lambda_0$ are the respective wavelengths. The redshift is defined as $z = (\lambda_0 - \lambda_e)/\lambda_e$, i.e., $(1+z) = \lambda_0/\lambda_e$. Therefore, for the tired light

$$d_p = \frac{1}{K}\ln(1+z). \tag{33}$$

It must reduce to Hubble's law $d = cz/H_0$ in the limit of $z \ll 1$, yielding $K = H_0/c$. Thus, the proper distance in the tired light (TL) model is given by

$$d_p = \frac{c}{H_0} \ln(1+z). \tag{34}$$

*Hybrid Models:* Since the distance traveled is the same in any model, the parameters of a hybrid tired light and expanding universe model are obtained by equating $d_p$ for the two. Thus for the hybrid CCC+TL model, we equate Eqs. (30) and (34), and write (with subscript $t$ for tired light and $c$ for the CCC model)

$$c \int_0^{z_c} \frac{dz}{(H_c+\alpha)(1+z)^{(3/2)} f(z)^{-(1/2)} - \alpha} = \frac{c}{H_t} \ln(1+z_t). \tag{35}$$

We will now determine $H_c$ and $H_t$ by considering the above expression in the limit of very low redshifts ($z \ll 1$) and keeping the terms with up to second order in $z$. Since the left-hand side involves an integral, we need to retain only up to first-order terms in the integrand.

Let us first consider $f(z)$. It is not practical to determine it from Eq. (27) in the small redshift limit $z = 0$. We go back to Eq. (24) and rewrite it with $a = 1/(1+z)$:

$$\frac{dz}{dt} = \alpha(1+z) - (H_0 + \alpha) f^{-(1/2)} (1+z)^{5/2}. \tag{36}$$

Since $f^{-(1/2)}$ varies slowly as compared to $(1+z)^{5/2}$, we can assume it is a constant equal to unity (i.e., its value at $z = 0$) in the above equation for solving it. With boundary condition $t = t_0$ at $z = 0$, the solution in the limit of $z \ll 1$ up to the first term in $z$ can be easily determined:

$$\frac{dz}{dt} = \alpha(1+z) - (H_0+\alpha)\left(1+\tfrac{5}{2}z\right) = -H_0 - \tfrac{1}{2}(5H_0+3\alpha)z \equiv A + Bz, \tag{37}$$

with solution

$$\exp[B(t-t_0)] = \left(1+\tfrac{B}{A}z\right) \Rightarrow (\exp[\alpha(t-t_0)])^{B/\alpha} = \left(1+\tfrac{B}{A}z\right) \Rightarrow f = \left(1+\tfrac{B}{A}z\right)^{\alpha/B}. \tag{38}$$

Since $A = -H_c$ for the CCC case, the above expression for $f$ reduces to

$$f(z) = \left(1 - \tfrac{\alpha}{H_c} z\right). \tag{39}$$

The left-hand side of Eq. (35):

$$c \int_0^{z_c} dz \left[(H_c+\alpha)(1+z)^{(3/2)}\left(1-\tfrac{\alpha z}{H_c}\right)^{-(1/2)} - \alpha\right]^{-1} = c\int_0^{z_c} dz\left[(H_c+\alpha)\left(1+\tfrac{3}{2}z\right)\left(1+\tfrac{\alpha z}{2H_c}\right) - \alpha\right]^{-1}$$

$$= c\int_0^{z_c} dz\left[H_c + (H_c+\alpha)\left(\tfrac{3}{2}+\tfrac{\alpha}{2H_c}\right)z\right]^{-1} = \frac{c}{H_c}\int_0^{z_c} dz\left[1 - \tfrac{(H_c+\alpha)}{2H_c}\left(3+\tfrac{\alpha}{H_c}\right)z\right] = \frac{cz_c}{H_c}\left(1 - \tfrac{(H_c+\alpha)}{4H_c}\left(3+\tfrac{\alpha}{H_c}\right)z_c\right). \tag{40}$$

Thus, by expanding the right-hand side, we may write Eq. (35) as

$$\frac{cz_c}{H_c}\left(1 - \frac{(H_c+\alpha)}{4H_c}\left(3+\frac{\alpha}{H_c}\right)z_c\right) = \frac{cz_t}{H_t}\left(1 - \frac{z_t}{2}\right). \tag{41}$$

Equating the first term on both sides of this equation and comparing it with Hubble's law yields $z_c/H_c = z_t/H_t = z/H_0$. Since the cumulative redshift $z$ obeys $(1+z) = (1+z_c)(1+z_t)$, in the small redshift limit, we get $z = z_c + z_t$. Therefore,

$$\frac{z_c}{H_c} = \frac{z_t}{H_t} = \frac{z}{H_0} = \frac{z_c+z_t}{H_0} \Rightarrow H_0 = H_c \frac{(z_c+z_t)}{z_c} = H_c\left(1+\frac{z_t}{z_c}\right) = H_c\left(1+\frac{H_t}{H_c}\right) = H_c + H_t. \tag{42}$$

Equating the second term on both sides of Eq. (41)

$$\frac{(H_c+\alpha)}{4H_c}\left(3+\frac{\alpha}{H_c}\right)z_c = \frac{z_t}{2} = z_c\frac{H_t}{2H_c} \Rightarrow H_t = \frac{(H_c+\alpha)}{2}\left(3+\frac{\alpha}{H_c}\right). \tag{43}$$

Let us now consider Eq. (35). With the above findings, we may write

$$c\int_0^{z_c}\frac{dz}{(H_c+\alpha)(1+z)^{(3/2)}f(z)^{-(1/2)}-\alpha} = c\left[\frac{(H_c+\alpha)}{2}\left(3+\frac{\alpha}{H_c}\right)\right]^{-1}\ln\left[\frac{1+z}{1+z_c}\right]. \tag{44}$$

Given $H_c$ and $\alpha$, this equation determines $(1+z_c)$ for any $(1+z)$ value and, therefore, also $(1+z_t)$. We can now use either of Eq. (31) or Eq. (34) to determine the proper distance $d_p$ while replacing $H_0$ and $z$ in there with $H_c$ and $z_c$ or $H_t$ and $z_t$ as applicable. When determining the luminosity distance, only $(1+z_c)$ is involved in time dilation, not $(1+z_t)$. The luminosity distance in the hybrid model is then $d_L = d_p(1+z_c)(1+z_t)^{1/2}$. Since $(1+z_t) = (1+z)/(1+z_c)$, the distance modulus for a hybrid model comprising the tired light and the CCC model becomes

$$\mu = 5\log_{10}(d_p/1\text{Mpc}) + 2.5\log_{10}[(1+z)(1+z_c)] + 25. \tag{45}$$

We can follow a similar approach for the ΛCDM+TL model (with subscript $l$ for the ΛCDM model). Equating Eq. (31) and Eq. (34)

$$\frac{c}{H_l}\int_0^{z_l}\frac{dz}{\sqrt{\Omega_{m,0}(1+z)^3+1-\Omega_{m,0}}} = \frac{c}{H_t}\ln(1+z_t). \tag{46}$$

In the limit of $z_l \ll 1$ and $z_t \ll 1$, and keeping the terms up to second order in redshift, we get the same results as for the CCC model by comparing the first-order terms but $H_t = 1.5\Omega_{m,0}H_l$ by comparing the second-order terms, and Eq. (44) is replaced with

$$\frac{c}{H_l}\int_0^{z_l}\frac{dz}{\sqrt{\Omega_{m,0}(1+z)^3+1-\Omega_{m,0}}} = \frac{c}{1.5\Omega_{m,0}H_l}\ln(1+z_t). \tag{47}$$

### III. RESULTS

*Fitting Pantheon+ Data:* We will now attempt fitting the Pantheon+ data (Scolnic et al. 2022, Brout et al. 2022) for supernovae type 1a (SNe1a). We have considered the following six models in matter dominated Euclidean Universe:
1. The ΛCDM model.
2. The CCC model.
3. The hybrid model, comprising the ΛCDM model and the tired light model labeled ΛCDM+TL.
4. The hybrid model, comprising the CCC model and the tired light model labeled CCC+TL.
5. The tired light model TL, modified to include an unknown flux loss correction term, labeled TL+, added to the luminosity distance, i.e., $5\log(1+z)^\beta$, with $\beta$ a data-fit parameter.
6. The standard tired light model, TL, with luminosity distance corrected only for the photon energy loss since the time dilation is not relevant for it.

We used the standard Curve Fitting tool in Matlab for this purpose by minimizing $\chi^2$ (Gupta 2020). The first five have two free parameters; thus, the degrees of freedom are identical. They are all within the error bars of the data (Fig. 1; only the first four model curves are plotted). The $\chi^2$ values of the first four models are within 0.1% of each other at 745.7 as its mean value (Table 1). Even the TL+ model's $\chi^2$ value is within 0.6% of the mean value. However, as could be expected, the $\chi^2$ value for the standard TL model is significantly larger at 2580. Thus, the Pantheon+ data rejects only the TL model but none of the others.

It should be mentioned that we have not taken into consideration the systematic uncertainties associated with the data. However, they would affect the fitting of all the models and $\chi^2$ values similarly and thus are not expected to impact the comparative study of the models.

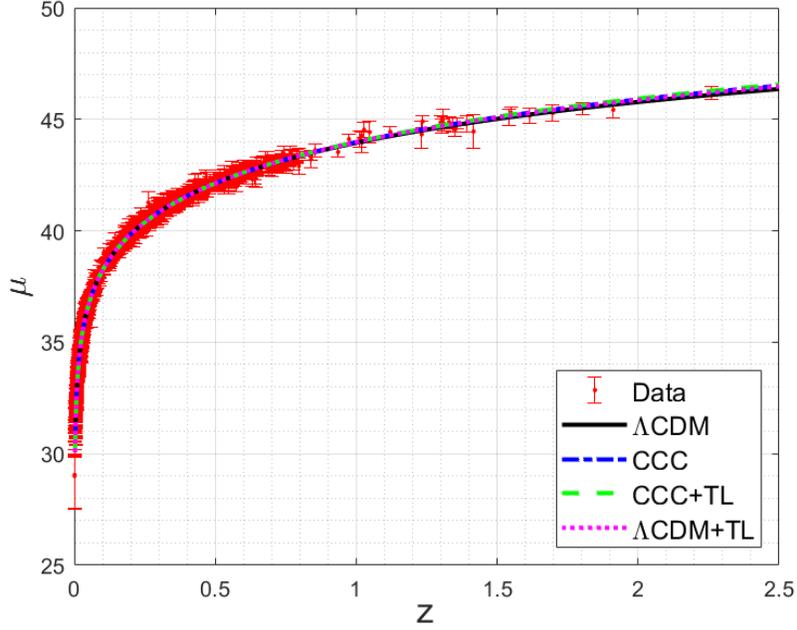

Figure 1 Pantheon+ (Scolnic et al. 2022, Brout et al. 2022) data fit for the four models discussed in the text.

The models' fit parameters and $\chi^2$ values, along with the deceleration parameter and the age of the universe, are presented in Table 1. It is relevant at this point to refer to the work of Lopez-Corredoia and Calvo-Torel (2022) wherein they compare the fits of several alternative models without dark energy by fitting Pantheon data (Scolnic et al. 2018).

TABLE 1. Pantheon+ data fit parameters, $\chi^2$ values, the deceleration parameter, and the age of the universe.

| Parameter | Unit | LCDM | CCC | LCDM+TL | CCC+TL | TL+ | TL |
|---|---|---|---|---|---|---|---|
| $H_0$ | Km s$^{-1}$ Mpc$^{-1}$ | 72.99 ± 0.34 | 72.70 ± 0.34 | 72.74 | 72.62 | 72.46 ± 0.34 | 64.42 ± 0.36 |
| $H_x$ | Km s$^{-1}$ Mpc$^{-1}$ | 72.99 | 72.7 | 60.48 ± 1.06 | 59.51 ± 1.06 | 0 | 0 |
| $H_t$ | Km s$^{-1}$ Mpc$^{-1}$ | 0 | 0 | 12.26 | 13.11 | 72.46 | 64.42 |
| $\Omega_{m,0}$ | NA | 0.3508 ± 0.0243 | NA | 0.1351 ± 0.0109 | NA | NA | NA |
| $\alpha/H_x$ | NA | NA | -0.4953 ± 0.0246 | NA | -0.7997 ± 0.0143 | NA | NA |
| $\beta$ | NA | NA | NA | NA | NA | 0.6418 ± 0.0196 | NA |
| $\chi^2$ | NA | 745.4 | 745.6 | 745.3 | 746.5 | 749.9 | 2580 |
| Fit Params. | NA | 2 | 2 | 2 | 2 | 2 | 1 |
| DOF | NA | 1699 | 1699 | 1699 | 1699 | 1699 | 1700 |
| $q_0$ | NA | -0.474 | -0.371 | -0.797 | -0.780 | NA | NA |
| $t_0$ | Gyr | 13.75 | 13.69 | 19.25 | 26.70 | NA | NA |

Unlike in a previous paper (Gupta 2020), wherein the absolute magnitude $M_B = -19.35$ was used to convert the apparent magnitudes reported in the Pantheon data (Scolnic 2018) to the distance moduli $\mu$, Pantheon+ data (Scolnic 2022) includes distance modulus, apparently based on $M_B = -19.253$. Since $H_0$ is degenerate with $M_B$, the choice of $M_B$ affects the determination of $H_0$ for different models. However, all the models yield about the same value of $H_0$. Thus, we do not expect the choice of absolute magnitude to affect the findings of this paper.

*Angular-Diameter Angle and Distance*: These are essential for testing a model based on the data on the size of galaxies at high redshift, especially those recently observed with the James Webb Space Telescope and discussed extensively in the literature (e.g., Lovyagin et al. 2022 and references therein).

The angular diameter distance $d_A$ is defined in terms of the physical size $\delta l$ of an object and its observed angular size $\delta\theta$ as $d_A = \delta l / \delta\theta$. Using the metric (Eq. 2), the object at a location $(r, \phi)$, i.e., $dr = 0$ and $d\phi = 0$, at time $t$ has a size given by

$$ds^2 = a(t)^2 f(t)^2 r^2 d\theta^2 \Rightarrow ds = a(t)f(t)r\delta\theta = \delta l \tag{48}$$

Therefore, with $r$ as the proper distance $d_p$, the angular diameter distance becomes

$$d_A = a(t)f(t)d_p. \tag{49}$$

We have to be cogent of the fact that the scale factor $a$ in a hybrid model relates to the redshift $z_x$ due to expanding universe only, i.e., $a = 1/(1 + z_x)$, whereas an observer measures the total redshift $z$. The function $f(t)$, of course, has no significance for the non-CCC models. In addition, $a(t)$ is irrelevant for the tired light model. Thus, when we know $d_p$ for a model, we can immediately compute $d_A$, and, therefore, the size of an object from its observed angular size.

Plots of angular-diameter distance for the five models are shown in Figure 2. Two of the plots jump out: i) the TL+ plot has no maximum, but that can be expected as no expansion of the universe is involved, and ii) the CCC+TL plot that has its peak value about four times higher than the ΛCDM value, and that too at $z > 10$. This dramatically affects the observed $\delta\theta$ and brightness of the distant galaxies. The plots are shown in Fig. 3 for 10 kpc objects against the backdrop of measured angular sizes of galaxies from multiple sources (the same as cited by Lovyagin et al. 2022; those points without errors in the original data are arbitrarily allowed a 5% error), including the latest JWST data. While all models are satisfactory at low redshifts, only TL+ and CCC+TL are acceptable for high redshifts.

It should be mentioned that the size evolution of galaxies, especially at z > 10, is not well measured. In particular, the measured angular sizes are subject to the distribution of stellar populations within galaxies and their light or mass profiles (e.g. outskirts of galaxies are likely too faint to be detected). Also, we have not addressed the uncertainties related to baryonic processes and feedback mechanisms that may potentially effective in shaping galaxy sizes of observed galaxies. Additionally, not all redshifts determined initially via photometric color selection, have been confirmed spectroscopically; some have been found spectroscopically at much lower redshift, e.g., a bright interloper at $z_{spec} = 4.91$ that was claimed as a photometric candidate at z ~ 16 (Harikane et al. 2023).

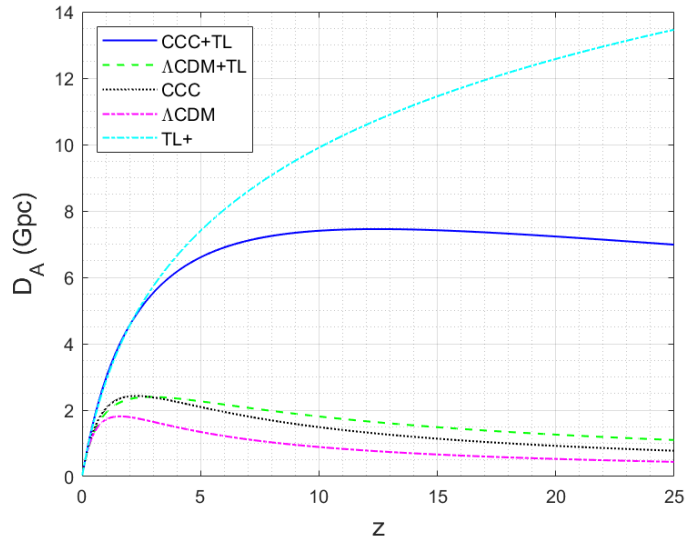

Figure 2. Angular diameter distance according to different models plotted against observed redshift.

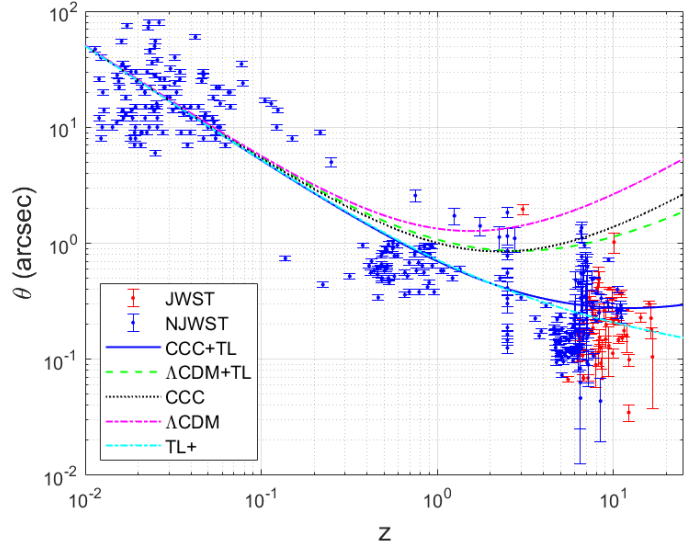

Figure 3. The angular size of 10 kpc objects for the five models against the backdrop of measured angular sizes of galaxies from multiple sources, including the latest JWST data and some pre-JWST data labeled as NJWST (provided by Lovyagin et al. 2022).

*Age of The Universe and Redshift*: Galaxies at high redshifts, especially those observed by JWST, appear as evolved and as massive as those at lower redshifts. How can this be possible, considering that the age of the universe at very high redshifts was less than half a billion years as per the ΛCDM model? This subject has been extensively discussed in the literature (e.g., Adams et al. 2022, Atek et al. 2022, Chen et al. 2022, Donnan et al. 2022, Finkelstein et al. 2022, Naidu et al. 2022a and 2022b, Ono et al. 2022, Tacchella et al. 2022, Wu et al. 2022, Yang et al. 2022, Labbe et al. 2023), so we will not delve on it. Nevertheless, we will try to see how the Universe ages under different models.

Let us first consider the CCC model. As per Eq. (24), we may write

$$\frac{da}{dt} = -\alpha a + (H_c + \alpha) f^{-(1/2)} a^{-(1/2)} \Rightarrow t = \int_0^{a_c} \frac{da}{-\alpha a + (H_c + \alpha) f^{-(1/2)} a^{-(1/2)}}. \tag{50}$$

Here $a_c = 1$ for the age of the universe today at $z_c = 0$. To find the age for a specific $z_c$, we replace $1/(1 + z_c)$ with $a_c$ and integrate. For non-hybrid models, $z$ replaces $z_c$.

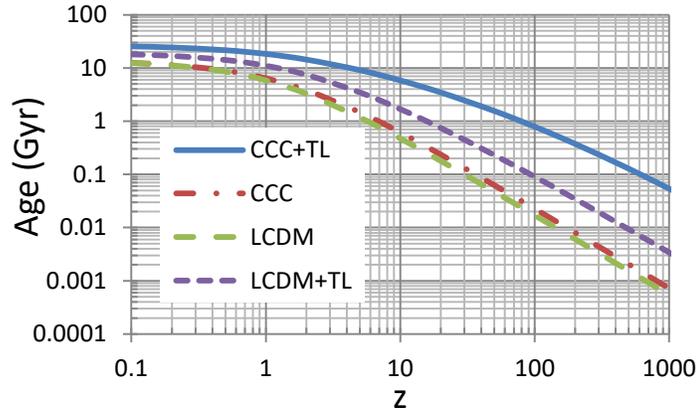

Figure 4. Age as a function of the observed redshift for the four models.

When working with a CCC+TL model, we have to find first the expanding universe component $z_c$ of $z$ as described under Eq. (44), and then use Eq. (50) to determine the age corresponding to the observed $z$. Similarly, we determine the ages for the ΛCDM and ΛCDM+TL models. The Universe age has no meaning for the TL and TL+ models.

Figure 4 shows how the age of the universe decreases with increasing redshift for the four models. Figure 5 shows the age increment with redshift for the two hybrid models compared to the currently expected age for the standard ΛCDM model. Both the hybrid models show significant increases, but the CCC+TL model provides a 10 to 20-fold increase at redshifts 10 to 20, giving enough time (5.8 Gyr at $z = 10$ and 3.5 Gyr at $z = 20$) for large, massive galaxies to form. In contrast, for the ΛCDM+TL model, corresponding age increases are 1.7 Gyr and 0.7 Gyr.

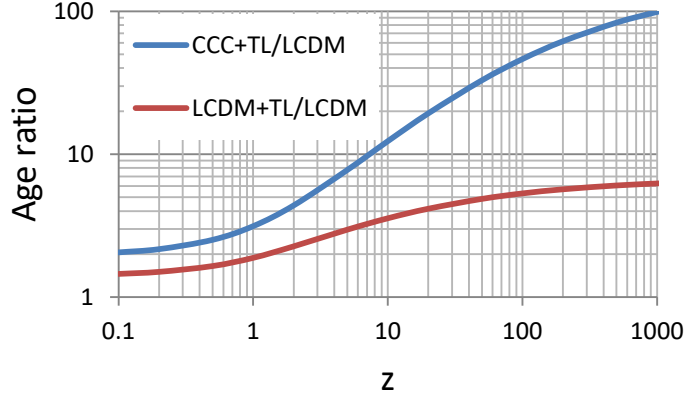

Figure 5. Age advantage of the two hybrid models over the ΛCDM model

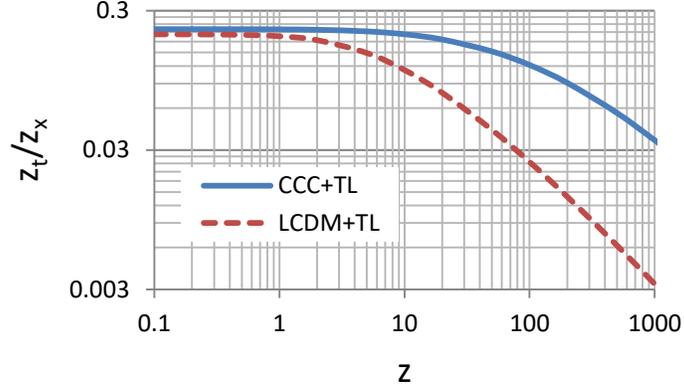

Figure 6. Tired light redshift as a fraction of the expanding Universe redshift.

*Tired Light Contribution to Redshift:* We were curious to know how the ratio of the tired light and expanding universe redshift evolves with the total redshift for the hybrid models. It is plotted in Figure 6. For the CCC+TL model, the tired light fraction starts at 22% at $z \approx 0$ and declines slowly to 20% at $z = 10$, to 18% at $z = 20$, and to 3% at $z = 1000$. For the ΛCDM+TL model, the corresponding values are 20%, 11%, 7%, and 0.3%. Despite their relatively small contribution, the tired light modifies the observed redshift considerably since redshifts are multiplicative through the relation $(1 + z) = (1 + z_x)(1 + z_t)$.

*Deceleration Parameter:* We will now calculate the deceleration parameter for the models. It is defined by the relation $q_0 = -\ddot{a}/(aH_0^2)$, $H_0$ being the expanding Universe part of the Hubble constant at the current time. Let us first consider the CCC model. From the first Friedmann equation (Eq. 11) for a flat Universe

$$H_0^2 = \frac{8\pi G}{3c^2}\varepsilon_0 - (\alpha^2 + 2\alpha H_0) = \frac{4\pi G}{3c^2}\varepsilon_0 = \left(H_0^2 + (\alpha^2 + 2\alpha H_0)\right)/2. \tag{51}$$

From the second Friedmann equation (Eq. 12) in the matter-dominated universe, using the above equation

$$\left(\frac{\ddot{a}}{a}\right)_0 = -\frac{4\pi G}{3c^2}\varepsilon_0 - \alpha H_0 = -\frac{1}{2}\left(H_0^2 + (\alpha^2 + 2\alpha H_0)\right) - \alpha H_0 \Rightarrow q_0 = \frac{1}{2} + 2\frac{\alpha}{H_0} + \frac{1}{2}\left(\frac{\alpha}{H_0}\right)^2. \quad (52)$$

Similarly, $q_0$ for the ΛCDM model may be expressed (Ryden 2017), since $\Omega_{\Lambda,0} = 1 - \Omega_{m,0}$, as

$$q_0 = \frac{1}{2}\Omega_{m,0} - \Omega_{\Lambda,0} = -1 + 1.5\Omega_{m,0}. \quad (53)$$

In the above expressions, we must use the expanding Universe component of the Hubble constant when applying them to the hybrid models. The deceleration parameters and the universe's ages for different models are shown in Table 1.

### IV. DISCUSSION

The main objective of this paper is to explore if the hybrid models that include tired light cosmology can explain the deep space high redshift observation of JWST on the large-scale structures of the universe, that is, the structure and evolution of galaxies. These observations show that the very early Universe galaxies were almost as bright, massive, and structurally as evolved as the galaxies in the late universe but with rather small angular diameters. Under the standard ΛCDM model, the physical sizes of these galaxies turn out to be about 10% of their expected sizes. It is because, according to this model, the angular diameter distance of objects reaches a maximum of about ~1.7 Mpc at $z = 1.6$ and decreases at higher redshift leading to the angular diameter $\theta$ of a standard size object increasing with increasing $z$. Thus, if one observes a smaller $\theta$, it is interpreted as a smaller size object. So, if a model has no angular diameter distance maximum, such as for the tired light models, or if the maximum is shifted to higher redshifts and thus has a higher value, then observed smaller $\theta$ will translate into larger object sizes. Observed over density of galaxies in the region of $z \sim 10$ (e.g. Whitler et al. 2023a, 2023b) can also be explained accordingly.

However, before any model can be considered seriously, it must pass the most basic test fitting the supernovae type 1a data compiled as Pantheon+ by Scolnic et al. (2022). Additionally, it should, in principle, not conflict with the observed isotropy of the CMB radiation. The standard tired light model TL does not pass the basic test, as is evident from its $\chi^2$ value of 2580 against all others less than 750 in Table 1. The modified tired light model TL+ with added unknown flux loss term $5\log(1+z)^\beta$ in the distance modulus equation gives a good fit to Pantheon+ with a $\chi^2$ value of 749.9 when $\beta = 0.6418 \pm 0.0196$. The inclusion of the additional term indicates that the TL model is missing something and that missing something is possibly the expanding universe component. However, neither of the tired light models can explain the extreme isotropy of the CMB radiation. Nevertheless, we have included the TL+ model in some of the comparative figures for discussion purposes. All the remaining models provide similarly great fits to Pantheon+ data (Fig. 1 and Table 1) and involve expansion of the universe and thus are compliant with CMB isotropy.

One may be concerned about how the peak luminosities of supernovae type 1a in the Pantheon+ data will be affected due to the variation of $G$ and other constants. The peak luminosity can be related to the Chandrasekhar mass $M_{ch}$ (Arnett 1982, Garcia-Berro et al 1999, Gaztañaga et al. 2001, Wright & Li 2018) with $M_{ch} \sim (hc/G)^{3/2}$. Since $h \sim f^2$, $c \sim f$, and $G \sim f^3$, we find that $M_{ch} \sim f^0$. Therefore, the coupling constants' variation does not affect the peak supernovae luminosity in the CCC model[1].

Coming back to angular diameter distance, its peak value of 7.45 Gpc at $z = 12.5$ is the highest among the four models of interest, and its profile is closest to the TL+ profile (Fig. 2). Thus, a small observed angular size $\theta$ of a galaxy does not translate into a smaller galaxy size. This is depicted in Fig. 3, which shows how one would observe $\theta$ of a 10 kpc object with increasing redshift under different models; the curves are superimposed over the $\theta$ of many galaxies from JWST observations and some pre-JWST observations. We notice that the CCC+TL model curve is very close to the TL+ curve, whereas the ΛCDM+TL model curve is not. This may mean that the ΛCDM+TL model (also, the CCC and ΛCDM models) do not faithfully represent the early universe evolution.

An alternative way to analyse the angular size data of the galaxies would be to see what each model yields as their physical size. Since the physical size is directly proportional to angular diameter distance, the size increase in different models compared to the ΛCDM model can be easily determined, as shown in Fig. 7. We see an insignificant difference using different models at redshifts less than 1. The difference becomes perceptible at higher and higher redshift. The multiplier for the CCC+Tl model becoming 5.6 at $z = 5$, 9.5 at $z = 10$, 12.8 at $z = 15$, and so on. Astronomers became concerned about the galaxy sizes using the ΛCDM model only at $z \geq 5$, especially at $z \geq 10$. Consequently, they suggested models and their improvements to create large structures of the universe in a shorter and shorter time (e.g., Adams et al. 2022, Atek et al. 2022, Chen et al. 2022, Finkelstein et al. 2022, Naidu et al. 2022, Ono et al. 2022, Tacchella et al. 2022, Wu et al. 2022, Yang et al. 2022, Labbe et al. 2023, Yung et al. 2023, Chantavat et al. 2023). Instead of attempting to *compress* the cosmic timeline for creating well-evolved massive galaxies in a very young universe to resolve the impossible early galaxy problem with the ΛCDM model (Melia 2023) and cosmological simulations (McCaffrey et al. 2023), it might be prudent to consider alternative models that *stretch* the timeline. An alternative scenario in support of the standard model has been proposed by Prada et al. (2023) by contending that during the early epochs of the universe the stellar mass-to-light ratio could not have reached the values reported by Labbe et al. (2023). However, one would need to explain the existence of very high $z$ quasars, such as those considered by Latif et al. (2023), and remarkably high IR luminosity of massive galaxies at $z \sim 8$ discovered by JWST (Akins et al. 2023).

---

[1] In an earlier paper (Gupta R.P., 2022, Mon. Not. R. Astron. Soc. 511, 4238.), the Planck constant scaling was incorrectly taken as $h \sim f$ rather than $h \sim f^2$.

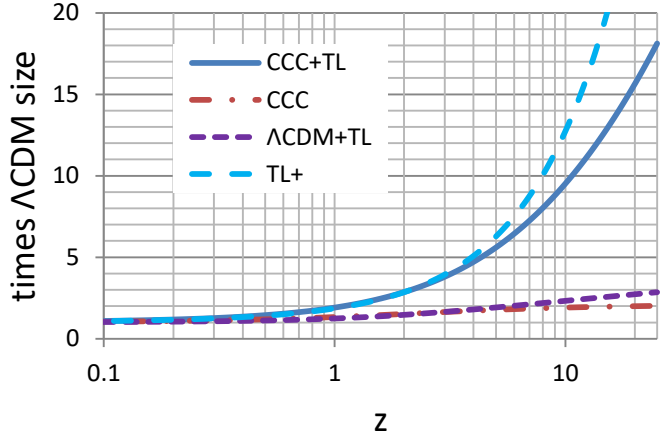

Figure 7. Object size under different models as a multiple of its size using the ΛCDM model.

We have shown in Fig. 4 how the cosmic timeline is stretched in the two hybrid models. We show the age of the universe under the two hybrid models relative to the age under the ΛCDM model. While the stretch is modest (1.40 times for the ΛCDM+TL model vs. 1.94 times for the CCC+TL model) at the current epoch, it becomes very significant at high redshift (2.97, 3.56, and 4.16 times at $z = 5$, 10, and 20, respectively, for the ΛCDM+TL model, and 7.76, 12.3, and 19.3 times for the same redshifts, respectively, for the CCC+TL model). The cosmic time stretch is thus up to 5.8 Gyr at $z = 10$ and 3.5 Gyr at $z = 20$; there is no need to invent new physics for the rapid formation of galaxies.

It would appear that except for the CCC+TL model, none of the models are able to reasonably resolve the impossible early galaxy problem; the TL models are unrealistic as they cannot reproduce the isotropy of CMB, and others do not provide realistic sizes of the early galaxies. Thus, the only realistic option is the CCC+TL hybrid model.

The CCC model is derived from Dirac's hypothesis of varying physical constants (Dirac 1937). His analysis was based on his large number theory and predicted the variation of $G$ and the fine structure constant $\alpha$. As discussed by Uzan (2011), if one constant varies, then others also must vary, and their variations correlated. From the local energy conservation laws, Gupta (2022) found that the variation of $c$, $G$, the Planck constant $h$, and the Boltzmann constant $k$ must follow $G \sim c^3 \sim h^{1.5} \sim k^{1.5}$ when the distance is measured with $c$. This leads to $\dot{G}/G = 3\,\dot{c}/c = 1.5\,\dot{h}/h = 1.5\,\dot{k}/k$, and therefore their variation can be represented by a single dimensionless function $f(t)$, i.e., *if one constant varies, then all of them do*. The standard procedure is to consider a power law function or a linear function. However, we found the most convenient form to use is $f(t) = \exp[\alpha(t - t_0)]$ with $\alpha$ determined by fitting the Pantheon+ data as $\alpha = -0.66H_0$ $(= \dot{c}/c)$; $\alpha$ may be considered to represent the strength of the coupling constants' variation. Since $\dot{G}/G = 3\,\dot{c}/c$, we get $\dot{G}/G = -2.0H_0$, about the same predicted by Dirac from his large number hypothesis. While there have been multiple attempts to constrain $|\dot{G}/G|$ by various methods to several orders of magnitude lower values (e.g.,Teller 1948; Morrison 1973; Sisterna & Vucetich 1990; Degl'Innocenti et al. 1995; Thorsett 1996; Corsico et al. 2013; Sahini & Shtanov 2014; Ooba et al. 2017; Hofmann & Müller 2018;

Genova et al. 2018; Wright & Li 2018; Bellinger & Christensen-Dalsgaard 2019; Zhu et al. 2019; Alvey et al. 2020; Vijaykumar et al. 2021) they all consider other constants to be pegged to their current value. However, keeping any of the constants fixed automatically forces $\alpha = 0$, and therefore $\dot{G}/G = 0$.

We believe the two-parameter $(H_0, \alpha)$ CCC model is possibly an extension of the $\Lambda$CDM model. Examining the Friedmann equations, one can imagine the CCC model as the $\Lambda$CDM model with a dynamic cosmological constant: (i) $\Lambda \rightarrow \Lambda(t) = -3(\alpha^2 + 2\alpha \dot{a}/a)$, consisting of a static term and a dynamic term in the first Friedmann equation, and (ii) only the dynamic term in the second Friedmann equation. In the CCC model, recalling that $\alpha$ numerically turns out to be negative, it replaces $\Lambda$ as the source of Universe expansion. One can even define energy density corresponding to $\alpha^2 + 2\alpha \dot{a}/a$. However, it remains to be seen how well the CCC model can fit the Planck data and explain other astrophysical and cosmological observations.

It should be mentioned that by treating the $g_{00}$ metric coefficient time dependent in the FLRW metric (Eq. 2), Gomide & Uehara (1981) were able to show that the local inertial effects are dependent on the overall structure of the Universe; cosmological models with positive curvature are Machian, whereas open ones are not. The CCC+TL model could, thus, be used to test the applicability of Mach's effect in the Universe.

It remains to be seen if the CCC+TL model can help address or explain other cosmology problems. Initially, we would want to determine if the new model is satisfactory for explaining CMB, BBN (big-bang nucleosynthesis), and BAO (baryonic acoustic oscillations) observations.

## V. CONCLUSION

JWST is perhaps playing the same role as HST did in the 1990s - reinventing cosmology. HST put the $\Lambda$CDM model on the pedestal. JWST is challenging standard $\Lambda$CDM. In this paper, we have attempted to show that an extension of the $\Lambda$CDM model with deemed dynamical cosmological constant, when hybridized with the tired light concept and parameterized with Pantheon+ data, provides a model, dubbed CCC+TL, that is compliant with the deep space observation of JWST on the angular sizes of high redshift galaxies. It stretches the cosmic time, especially at high redshifts, to allow the formation of large galaxies. It eliminates the need for stretching and tuning existing models to produce such structures in the early universe, thus amicably resolving the *impossible early galaxy* problem.

**Acknowledgment**

The author is grateful to Professor Nikita Lovyagin for sharing the data he and his collaborators used in their study; his work inspired the current work. He wishes to thank Macronix Research Corporation for the unconditional research grant for this work. He expresses his gratitude to the anonymous reviewer for constructive comments and suggestion for improving the quality and clarity of the paper.

**Data availability**

References have been provided for the data used in this work.

[i] Email: rgupta4@uottawa.ca